\documentclass[journal]{elsart5p}
\usepackage{graphicx}
\usepackage{subfigure}
\usepackage{epsfig}

\begin{document}


\begin{frontmatter}

\title{Monolithic Pixel Sensors in Deep-Submicron SOI Technology\\ with Analog and Digital Pixels}

\author[LBNL,UCB]{Marco Battaglia},
\author[INFN]{Dario Bisello},
\author[LBNL]{Devis Contarato\corauthref{cor1}},
\author[LBNL]{Peter Denes},
\author[LBNL,INFN]{Piero Giubilato},
\author[UCB]{Lindsay Glesener},
\author[INFN]{Serena Mattiazzo},
\author[LBNL]{Chinh Vu}
\address[LBNL]{Lawrence Berkeley National Laboratory, 
Berkeley, CA 94720, USA}
\address[UCB]{Department of Physics, University of California at 
Berkeley, CA 94720, USA}
\address[INFN]{Dipartimento di Fisica, Universit\`a di Padova and INFN,
Sezione di Padova, I-35131 Padova, Italy}
\corauth[cor1]{Corresponding author. Address: Lawrence Berkeley National Laboratory,
1 Cyclotron Road, Berkeley, CA 94720 (USA). E-mail: \tt{DContarato@lbl.gov} \rm (Devis Contarato).}

\begin{abstract}
This paper presents the design and test results of a prototype monolithic pixel 
sensor manufactured in deep-submicron fully-depleted Silicon-On-Insulator (SOI) 
CMOS technology. In the SOI technology, a thin layer of integrated electronics is 
insulated from a (high-resistivity) silicon substrate by a buried oxide. Vias
etched through the oxide allow to contact the substrate from the electronics 
layer, so that pixel implants can be created and a reverse bias can be applied.
The prototype chip, manufactured in OKI 0.15~$\mu$m SOI process, features both
analog and digital pixels on a 10~$\mu$m pitch. Results of tests performed with 
infrared laser and 1.35~GeV electrons and a first assessment of the effect of 
ionising and non-ionising doses are discussed. 
\end{abstract}

\begin{keyword}
Monolithic pixel sensor; SOI; CMOS technology; Particle detection
\end{keyword}

\end{frontmatter}

%
\section{Introduction}
Silicon on insulator (SOI) technology allows the fabrication of CMOS integrated
circuits on a thin silicon layer which is electrically insulated from the rest of
the silicon wafer by means of a buried-oxide (BOX). The small active volume and
low junction capacitance ensure latch-up immunity, low-threshold operation and
low noise, thus favouring high-speed, low power designs. The isolation of the
electronics from a high-resistivity substrate, together with the capability
of contacting the latter by means of vias through the BOX, allows the design
of monolithic pixel sensors for radiation detection which offer many advantages
with respect to devices fabricated in a standard bulk CMOS process. A full CMOS
circuitry can be integrated in each pixel, thus allowing for complex digital
designs, while the depletion of the sensor wafer results in an improved charge
collection efficiency. Furthermore, present-day deep-submicron processes allow
for a high miniaturisation and integration of complex architectures in devices
with small pixel pitch. A proof of principle of the concept was obtained by the
SUCIMA Collaboration, using a 3~$\mu$m process from IET,
Poland~\cite{marczewski2005,marczewski2006,niemec2006}. The recent availability of
deep-submicron, fully-depleted SOI CMOS process from OKI, Japan~\cite{oki}
opened up the possibility of fabricating SOI monolithic pixel sensors with
small pixel pitches and larger integration capabilities. The process features a
full CMOS circuitry implanted on a 40~nm thin Si layer on top of a 200~nm thick BOX.
The thickness of the CMOS layer is small enough for the layer to be fully depleted
at typical operational voltages. The sensor substrate is 350~$\mu$m thick and has
a resistivity of 700~$\Omega\cdot$cm. The adequacy of the OKI process for the
fabrication of monolithic pixel sensors was first demonstrated at KEK in
2006~\cite{tsuboyama2007}. A first prototype sensor designed at LBNL was manufactured 
in 2007 in the OKI 0.15~$\mu$m FD process. The chip features both analog and digital 
pixels. The first results on the detection of 1.35~GeV electrons with the analog pixels
have already been reported in~\cite{battaglia_nima}. In this paper we discuss further 
tests of the analog pixels to determine the charge sharing and the cluster position 
resolution using an IR laser and the first results of the characterisation of the 
digital pixels using 1.35~GeV electrons form the BTS beam at the LBNL Advanced Light 
Source (ALS).

%
\section{Chip Design, Readout and Data Analysis}
\begin{figure}[ht!]
\begin{center}
\epsfig{file=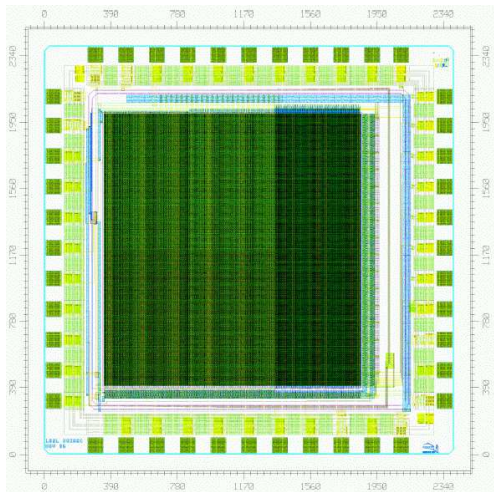,width=6.5cm}
\end{center}
\caption[]{Layout of the LBNL LDRD-SOI1 prototype pixel chip.}
\label{fig:chip_layout}
\end{figure}

The first prototype sensor, named LDRD-SOI-1, was manufactured in 2007 in the OKI 0.15~$\mu$m FD
process~\cite{battaglia_nima}. The chip features an array of 160$\times$150 pixels on a 10~$\mu$m
pitch, subdivided into two analog sections and one digital section, each comprising 160$\times$50
pixels (see Figure~\ref{fig:chip_layout}). 
The two analog sections implement a simple 3-transistor (3T) architecture based on 
thin-oxide 1.0~V transistors and thick-oxide 1.8~V transistors. The digital pixels operates at a 
bias V$_{\rm DD}$=1.8~V and implement an in-pixel latch triggered by a comparator connected to an
adjustable voltage threshold which is common to the whole matrix. No amplifier is present in the
digital pixels, so that static power dissipation is avoided, the pixel being active only when the
latch is triggered.

A potential limitation of the SOI technology comes from the back-gating effect. The reverse bias
of the silicon substrate increases the potential at the silicon surface, so that the buried oxide
acts as a second gate for the CMOS electronics on top, typically causing a shift in the transistor
thresholds as a function of the increasing depletion voltage. The effect could be evaluated from
single transistor test structures implemented at the chip periphery. Tests of complementary
$p$-type and $n$-type MOSFETs showed a threshold shift of $\sim$200~mV for a substrate bias
of 15~V. TCAD simulations performed with the Synopsys {\tt Taurus Device} package have  shown
that the inclusion of a floating $p$-type guard-ring around each pixel is beneficial in keeping
the field low in the area between diode implants, thus limiting potential back-gating effects
on the CMOS electronics on top of the buried oxide. A series of floating and grounded guard-rings
was also implemented around the pixel matrix and around the peripheral I/O electronics.

Each 8000-pixel analog section is read out independently by a custom FPGA-based
readout board with 14-bit ADCs. Pixels are clocked at 6.25~MHz, with an integration
time of 1.382~ms for the analog pixels, while the integration time for the digital
pixels is tunable. Correlated Double Sampling (CDS) is performed for the analog pixels
by acquiring two frames of data without resetting the pixels in-between the readings
and subtracting the first frame from the second. The binary output of the digital
pixels is simply buffered through the FPGA to the DAQ board digital outputs connected 
to a National Instruments PCI 6533 digital acquisition board installed on the PCI bus 
of a control PC. Data is processed on-line by a LabView-based program and converted 
into the {\tt lcio} format~\cite{Gaede:2003ip}. Offline data analysis is based on a 
set of dedicated processors developed in the {\tt Marlin} C++ reconstruction 
framework~\cite{Gaede:2006pj}, which perform data quality and event selection, 
clusterisation and hit position reconstruction.

\section{3T Analog Pixels}

The response of the analog sections has been first tested with a 1060~nm IR laser focused
to a $\simeq$20~$\mu$m spot, and measuring the pulse height in a 5$\times$5 pixel matrix,
centred around the laser spot centre. 
\begin{figure}[hb!]
\begin{center}
\epsfig{file=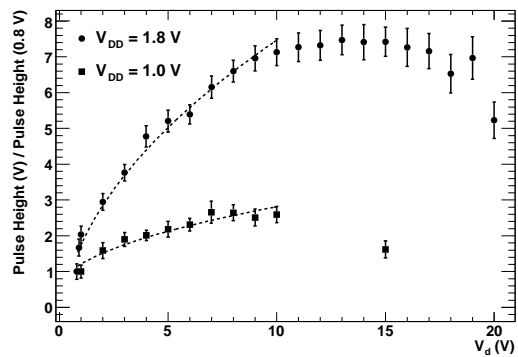,width=7.0cm}
\end{center}
\caption[]{Cluster pulse height normalised to that measured at 
$V_d$ = 0.8~V for a focused 1060~nm laser spot as a function of the 
substrate bias, $V_d$ (from~\cite{battaglia_nima}).}
\label{fig:laser}
\end{figure}
For depletion voltages $V_d<$10~V, the measured
signal was found to increase proportionally to $\sqrt{V_d}$, as expected from the increase
of the depletion region thickness. For larger $V_d$ values, the signal was found to first
saturate and then decrease for $V_d\ge$15~V, suggesting an effect due to the back-gating
of the MOSFETs in the pixel and/or in the output buffer. Such effect appears in the 1.0~V 
transistor pixels at lower $V_d$ values compared to the 1.8~V transistor
pixels (see Figure~\ref{fig:laser}).
\begin{figure}[h!]
\begin{center}
\epsfig{file=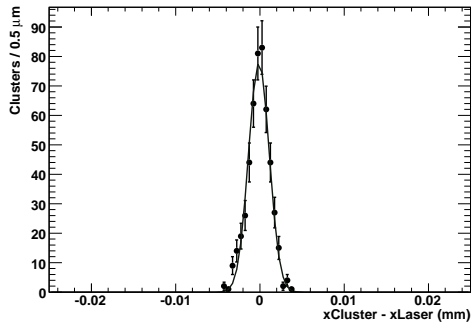,width=7.0cm}
\end{center}
\caption{Distribution of the differences between the position of a 5~$\mu$m laser spot 
and the reconstructed cluster position in the SOI analog pixels. The measurement has 
been performed with a S/N of 20 and a depletion voltage $V_d$ = 7~V. The r.m.s. of the 
fitted Gaussian function is 1.2~$\mu$m.}
\label{fig:res_sn20}
\end{figure}

The spatial resolution of the analog pixels has been determined by focusing the
1060~nm laser beam to a $\simeq$5~$\mu$m Gaussian spot and performing pixel scans
by shifting the spot along single pixel rows in steps of 1~$\mu$m using a stepping
motor with a positioning accuracy of 0.1~$\mu$m. The laser hit position is reconstructed
from the centre of gravity of the pixel cluster and the resolution is obtained from
the spread of the reconstructed cluster position for the events taken at each point
in the scan (see Figure~\ref{fig:res_sn20}). 
\begin{figure}[t]
\begin{center}
\epsfig{file=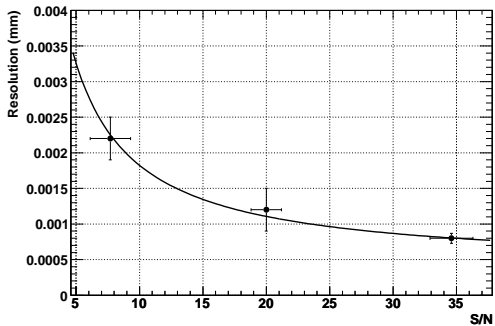,width=7.0cm}
\end{center}
\caption{Single point resolution as a function of the S/N ratio obtained from
scans of the analog pixels  performed with a 1060~nm laser
focused to a 5~$\mu$m spot, for a depletion voltage $V_{d}$=7~V. Data points represent
the mean of the distribution of the differences between the laser spot position
and reconstructed cluster position, the error bars give their r.m.s. values. 
The overlayed function shows the expected (S/N)$^{-1}$ scaling.}
\label{fig:res}
\end{figure}
The laser intensity is varied in order to obtain different S/N values,
from values below that observed for 1.35~GeV electron signals (see next section) up to 
S/N~$\simeq$~35. Results obtained for different S/N values and for a depletion voltage
$V_d$=7~V are shown in Figure~\ref{fig:res}: pixels with 10~$\mu$m pitch have a single
point resolution of 1~$\mu$m for a S/N ratio of 20 or larger and the measured resolution
scales as the inverse of the S/N, as expected. 

The response of the analog pixels to high-momentum charged particles has been tested on 
the 1.35~GeV electron beam extracted from the booster of the LBNL Advanced Light Source
(ALS) and results have been reported in detail in~\cite{battaglia_nima}. The pixel 
multiplicity in a cluster was found to slightly decrease with increasing depletion voltage, 
while the cluster pulse height increased up to 10~V. At 15~V the cluster signal and the 
efficiency of the chip decreased, as observed in the laser tests. A signal-to-noise
ratio up to 15 was measured with the 1.8~V analog section for 5~V$\le V_d \le$15~V.

\section{Digital Pixels}

The response of the digital pixels to high-momentum charged particles has also been tested 
on the ALS 1.35~GeV electron beam. The digital pixels are triggered directly by the 1~Hz 
booster extraction signal and are latched and read out after allowing an integration time 
of 10~$\mu$s. Data have been taken at different depletion voltages, up to 35~V. The depletion 
voltages used correspond to an estimated depletion thickness from 56~$\mu$m for $V_d$ = 15~V, 
up to 80~$\mu$m for $V_d$ = 30~V.

\begin{figure}[ht]
\begin{center}
\epsfig{file=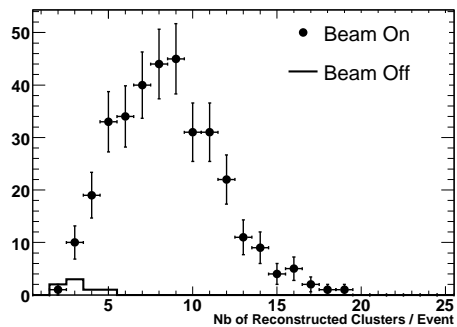,width=7.0cm}
\end{center}
\caption[]{Hit multiplicity for 1.35~GeV $e^-$s for the digital pixels at 
a depletion voltage of 30~V (markers with error bars).
The distribution of fake hits reconstructed in the absence of beam (continuous line)
is also shown for comparison.}
\label{fig:als_digi}
\end{figure}

For each run the first 100 events have been used to map the noisy pixels. 
Pixels giving a signal for more than 25 events have been flagged as noisy and 
masked for the rest of the run. Figure~\ref{fig:als_digi} shows the hit multiplicity 
observed in the digital pixels for events taken with and without beam, for 
$V_d$=30~V. A clear excess of hits can be seen in presence of beam. The small 
background obtained in the absence of beam is due to noisy pixels which survive 
the cluster reconstruction cuts applied in the data analysis. In order to evaluate
the effect of $V_d$ on the particle response, dedicated data runs have been taken 
at low beam intensity with a single detector reference plane, located 2~cm upstream 
from the LDRD-SOI1 sensor.  The reference plane consists of a 50~$\mu$m-thin MIMOSA-5 
CMOS pixel sensor. The MIMOSA-5 chip was developed by IPHC, Strasbourg 
(France)~\cite{mimosa, deptuch-ref} and features a 1.7$\times$1.7~cm$^2$ active area 
with more than 1~Million pixels on a 17~$\mu$m pitch, arranged in four independent sectors. 
The performance of thin MIMOSA-5 sensors, the readout system and analysis procedures 
adopted are presented in details in~\cite{Battaglia:2006tf,Battaglia:2008nj}.
The particle flux in the area corresponding to the active region of the LDRD-SOI1 
digital pixel sector has been monitored from the number of electron hits reconstructed 
on the reference plane, on a run by run basis. We compute the average number of 
hits per spill on the digital pixels, for different depletion voltages, and correct it 
for the relative change of the beam intensity, as determined by the reference plane. 
These changes are in the range of 5-40\%, for the runs of interest. Results are summarized 
in Table~\ref{tab:als_digi}. 
Signals from beam particles are observed on the digital section of the chip by applying 
a depletion voltage $\ge$~20~V, while the analog pixels stop functioning properly 
above $\simeq$~15~V. This can be explained by considering that the analog 
threshold of the in-pixel comparators is affected by back-gating, while the digital 
circuitry in each pixel is only active when triggered, i.e. for times much shorter 
compared to the analog pixels. At the same time, larger substrate voltages are needed 
to obtained large enough signals to be above threshold. These two effects seem to 
combine giving the best particle detection capabilities for $\simeq$~25~V. Using the 
particle flux reconstructed on the reference layer and its efficiency, as obtained from 
simulation~\cite{Battaglia:2008nj}, we estimate the efficiency of the SOI digital 
pixels to be of the order of 0.3 to 0.5 for 20~V$\le V_d\le$35~V.

\begin{table}[h!t!b!]
\begin{center}
\begin{tabular}{|c|c|c|c|}
\hline  \textbf{$V_d$} & \textbf{$\frac{Nb. Clusters}{Spill}$} & 
\textbf{$\frac{Nb. Clusters}{Spill}$} & \textbf{$<$Nb Pixels$>$} \\ 
\textbf{(V)}           & \textbf{beam on}                      
& \textbf{beam off}    &\textbf{in Cluster}       \\ 
\hline
20                     & 3.7$\pm$0.1         
& 0.02                 & 1.78                     \\ 
25                     & 5.3$\pm$0.1                                  
& 0.03                 & 1.32                     \\
30                     & 4.7$\pm$0.1                                  
& 0.03                 & 1.26                     \\
35                     & 4.2$\pm$0.1                                  
& 0.02                 & 1.14                     \\
\hline
\end{tabular}
\end{center}
\caption{Summary of beam test results obtained with 1.5~GeV electrons on the digital
pixels. The average number of clusters per beam spill recorded
with beam on and beam off and the average pixel multiplicity in a cluster are given for
different values of $V_d$. The number of cluster per spill is corrected by changes in the 
beam intensity, obtained from the reference plane.}
\label{tab:als_digi}
\end{table}
Data could be taken up to depletion voltages $V_d$=35~V, however with a much decreased counting 
rate. The average number of pixels in a cluster was found to decrease from 1.8 to 1.1 for $V_d$ 
increasing from 20~V to 35~V, consistently with an increased electric field in the detector 
substrate, as observed also for the analog pixels. 

\section{Radiation Hardness Tests}
The SOI technology is expected share the intrinsic radiation tolerance of deep-submicron electronics 
and the capability of sustaining significant non-ionising doses of high resistivity Si. However, the 
thick buried oxide may be sensitive to ionising doses, which lead to positive charge trapping and 
consequently to an increase of the back-gate potential. 

\begin{figure}[ht!]
\begin{center}
\epsfig{file=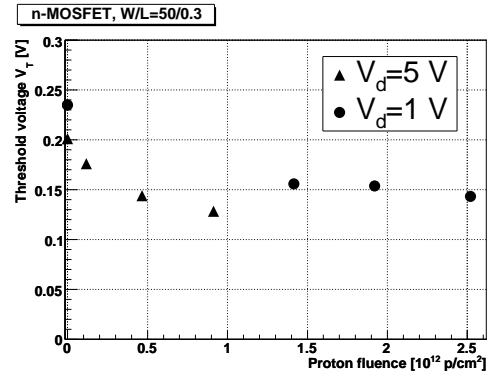,width=7.0cm} \\
\end{center}
\caption{Threshold voltage for a 1.0~V $n$-MOS test transistor with $W/L$=50/0.3 as a 
function of 30~MeV proton fluence for depletion voltages $V_d$=1~V and $V_d$=5~V.}
\label{fig:irradp}
\end{figure}

A preliminary assessment of the effect of ionising radiation on single 
transistors and of non-ionising radiation on the analog pixels has been obtained.
Irradiations were performed at the BASE Facility of the LBNL 88-inch
Cyclotron~\cite{cyclotron}. The first test was performed with 30~MeV protons on
single transistors. The chip was mounted on the beam behind a 1-inch diameter
collimator, and the terminals of two test transistors (one $p$-MOSFET and one
$n$-MOSFET) were connected to a semiconductor parameter analyser so that the
transistor characteristics could be measured in-between irradiation steps. During
the irradiation steps, the transistor terminals were kept grounded. The irradiation
was performed with a flux of $\sim$6$\times$10$^7$~p/cm$^2$s, up to a total fluence
of 2.5$\times$10$^{12}$~p/cm$^2$, corresponding to a total dose of $\sim$600~kRad.
Figure~\ref{fig:irradp} shows the variation in the threshold voltage for the $n$-MOS
test transistor as a function of the proton fluence. 
An initial substrate voltage
$V_d$=5~V was used, but after a fluence of about 1$\times$10$^{12}$~p/cm$^2$ the
transistor characteristics could not be properly measured, and a reduced substrate
bias of $V_d$=1~V needed to be applied in order to recover the transistor characteristics.
We interpret this effect as due to radiation-induced charge build-up in the buried oxide
which effectively increases back-gating. The total threshold variation is indeed
significant ($\sim$100~mV) also for a low substrate bias (i.e. $V_d$=1~V); the effect
is much larger than what would be expected at such doses from radiation damage in the
transistor thin gate oxide. Similar results were found on the $p$-MOS test transistor.
\begin{figure}[ht!]
\begin{center}
\epsfig{file=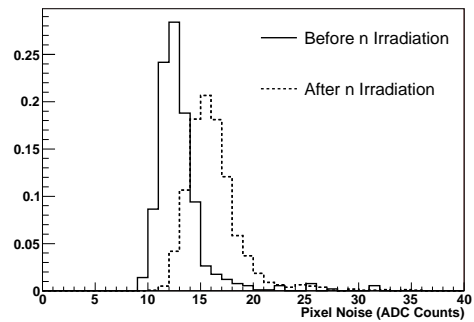,width=7.0cm} \\
\end{center}
\caption{Distribution of the noise of the analog pixels before (continuous line) and 
after (dashed line) irradiation with 1.2$\times$10$^{13}$~$n$~cm$^{-2}$ for $V_d$ = 10~V.
The histogram are normalised to unit area for comparison.}
\label{fig:irradn}
\end{figure}
The effect of non-ionising radiation has been studied exposing the LDRD-SOI1 
detector chip to neutrons produced from 20~MeV deuteron breakup on a 
thin target at the LBNL 88-inch cyclotron~\cite{mcmahan}. The detector chip 
was located 8~cm downstream from the target and activation foils were placed 
just behind it to monitor the fluence. Deuteron breakup produces neutrons on a 
continuum spectrum from $\sim$~1~MeV up to $\sim$~14~MeV. A beam current of 
800~nA was used, corresponding to an estimated flux at the chip position of 
$\simeq$~4~$\times$ 10$^{8}$~$n$~cm$^{-2}$~s$^{-1}$. The measured $\gamma$ 
activity on the foils corresponds to a total fluence of 
1.2$\times$10$^{13}$~$n$~cm$^{-2}$. The noise of the sensor has been measured 
before and after the irradiation as a function of the depletion voltage, $V_d$.
The detector is still functional after irradiation but we observe an increase of 
noise, which varies from +25~\%, for $V_d$ = 5~V, to +52~\%, for $V_d$ = 20~V, 
at room temperature (see Figure~\ref{fig:irradn}). 
We interpret this increase as due to leakage current. 
The pixel noise of the irradiated chip has therefore been measured, for 
$V_d$ = 10~V, as a function of the operational temperature, $T$, in the range 
-5~$^{\circ}$C $ < T <$ +20~$^{\circ}$C. At temperatures below +5~$^{\circ}$C, 
the noise level measured before irradiation is recovered.

\section{Conclusions and Outlook}
A prototype monolithic pixel sensor including both analog and digital pixel cells has 
been designed and produced in OKI SOI technology, which combines deep-submicron CMOS 
electronics and a high resistivity substrate in the same device. The response of the 
prototype chip has been studied using both focused laser beams and the 1.35~GeV $e^-$ beam 
at the LBNL ALS. Both the analog and digital pixels were found to be functional. A S/N ratio 
of 15 has been obtained with the electron beam on the analog pixels up to depletion voltages 
$V_d$=15~V, and a single point resolution of $\simeq$~1~$\mu$m has been estimated using 
measurements performed with a narrowly focused IR laser beam. The digital section of the
chip could be operated at higher depletion voltages compared to the analog part, as the 
digital circuitry in the pixels appears to be less affected by back-gating.
Total ionising dose tests performed on single transistor test structures hint at the build-up
of charge trapped in the buried oxide which enhance the effect of back-gating of the CMOS
electronics. A moderate increase of the pixel noise from exposure to  10$^{13}$~$n$~cm$^{-2}$ 
has been observed. These results establish the feasibility of monolithic pixel sensors in SOI 
technology, which is of great interest for the possibility to implement complex readout 
architectures combined with a high-resistivity, depleted substrate ensuring faster charge 
collection and larger signals, compared to bulk CMOS pixel sensors. Potential applications 
of this technology range from  vertex tracking in high-energy physics experiments to soft 
X-ray detection at synchrotron facilities and fast nano-imaging imaging for beam diagnostics
and monitoring. A second prototype chip, implementing digital pixels with in-pixel charge 
storage and correlated double sampling and fast readout has been recently fabricated 
in OKI 0.20~$\mu$m FD SOI process and is currently under test.

\section*{Acknowledgements}
This work was supported by the Director, Office of Science, of the U.S. Department
of Energy under Contract No. DE-AC02-05CH11231. We thank the staff of the LBNL Advanced 
Light Source and 88-inch Cyclotron for their assistance and for the excellent performance 
of the machines.


\begin{thebibliography}{99}

\bibitem{marczewski2005}
J.~Marczewski et al., Nucl.\ Instrum.\ Meth.\ A {\bf 549} (2005) 112. 

\bibitem{marczewski2006}
J.~Marczewski et al., Nucl.\ Instrum.\ Meth.\ A {\bf 560} (2006) 26. 

\bibitem{niemec2006}
H.~Niemiec et al., Nucl.\ Instrum.\ Meth.\ A {\bf 568} (2006) 153. 

\bibitem{oki}
OKI Electric Industry Co.~Ltd., Japan. \tt{http://www.oki.com}\rm

\bibitem{tsuboyama2007}
T.~Tsuboyama et al., Nucl.\ Instrum.\ Meth.\ A {\bf 582} (2007) 861. 

\bibitem{battaglia_nima}
M.~Battaglia et al., Nucl.\ Instrum.\ Meth.\ A {\bf 583} (2007) 526.
[arXiv:0709.4218 [physics.ins-det]].

\bibitem{Gaede:2003ip}
  F.~Gaede {\it et al.}
  in the {\it Proc. of 2003 Conf. for Computing in High-Energy and Nuclear Physics}
  (CHEP 03), La Jolla, California, 24-28 Mar 2003, pp TUKT001,
  [arXiv:physics/0306114].

\bibitem{Gaede:2006pj}
  F.~Gaede,
  Nucl.\ Instrum.\ Meth.\ A {\bf 559} (2006) 177.

\bibitem{mimosa}
 Yu.~Gornushkin {\it et al.}, Nucl.\ Instrum.\ and Meth.\ A {\bf 513} (2003) 291.

\bibitem{deptuch-ref}
 G.~Deptuch, Nucl.\ Instrum.\ and Meth.\ A {\bf 543} (2005) 537.

\bibitem{Battaglia:2006tf}
  M.~Battaglia {\it et al.},
  Nucl.\ Instrum.\ Meth.\  A {\bf 579} (2007) 675.

\bibitem{Battaglia:2008nj}
  M.~Battaglia {\it et al.},
  Nucl.\ Instrum.\ Meth.\  A {\bf 593} (2008) 292
  [arXiv:0805.1504 [physics.ins-det]].

\bibitem{cyclotron}
 M.A.~McMahan, Nucl.\ Instrum.\ Meth.\ B {\bf 241} (2005) 409

\bibitem{mcmahan}
 M.A.~McMahan, Nucl.\ Instrum.\ Meth.\  B {\bf 261} (2007), 974.

\end{thebibliography}
\end{document}